\documentclass[structabstract]{aa}
\usepackage{lineno}
\usepackage{graphicx}                  

\begin{document}

 \title{Dynamics of pebbles in the vicinity of a growing planetary
   embryo: hydro-dynamical simulations}

\author{\textbf{A. Morbidelli\inst{1} \and D. Nesvorny\inst{2}}} 

\institute{Dep. Lagrange, UNSA, CNRS, OCA, Nice, France, \and 
Dept. of Space Sudies, SwRI, Boulder,   CO, USA} 

\offprints{A. Morbidelli, \email{morby@oca.eu}} 

\abstract{Understanding the growth of the cores of giant planets is a
  difficult problem. Recently, Lambrechts \& Johansen (2012; LJ12)
  proposed a new model in which the cores grow by the accretion of
  pebble-size objects, as the latter drift towards the star due to gas
  drag.}{We investigate the dynamics of pebble-size objects in the
  vicinity of planetary embryos of 1 and 5 Earth masses and the
  resulting accretion rates.}{We use hydrodynamical simulations, in
  which the embryo influences the dynamics of the gas and the pebbles
  suffer gas drag according to the local gas density and
  velocities.}{The pebble dynamics in the vicinity of the planetary
  embryo is non-trivial, and that it changes significantly with the
  pebble size. Nevertheless, the accretion rate of the embryo that we
  measure is within an order of magnitude of the rate estimated in
  LJ12 and tends to their value with increasing pebble-size.}{The
  model by LJ12 has the potential to explain the rapid growth of giant
  planet cores. The actual accretion rates however, depend on the surface
  density of pebble size objects in the disk, which is unknown to date.}

\keywords{Accretion, Planets: formation, hydrodynamics}

\maketitle

\section{Introduction}

The formation of the massive cores of giant planets within the
short timescale allowed by the survival of a proto-planetary disk of
gas and solids (a few My; Haisch et al. 2001) is still an open
problem. In the classical view, these cores form by collisional
coagulation from a disk of km-sized planetesimals, through the
well-known processes of {\it runaway} (Greenberg et al. 1978;
Wetherill \& Stewart 1989) and {\it oligarchic} growth (Ida \&
Makino 1993; Kokubo \& Ida 1998). In principle these processes
should continue until the largest objects achieve an {\it isolation
  mass}, which is a substantial fraction of the initial total mass of
local solids. If the initial disk is sufficiently massive (about 10
times the so-called Minimal Mass Solar Nebula or MMSN;
Weidenschilling 1977; Hayashi 1981), 
it is expected that cores of $\sim 10$ Earth masses ($M_E$)
form beyond the so-called {\it
  snowline} (The orbital radius beyond which temperature is cold
enough that water condenses into ice; Podolak \& Zucker 2004), as
required in the core-accretion model for giant planet
formation (Thommes et al. 2003; Goldreich et al. 2004; Chambers 2006). 

$N$-body simulations, though, show that reality is not so
simple. When the cores achieve a mass of about 1~$M_\oplus$ they start
to scatter the planetesimals away from their neighborhood, instead of
accreting them (Ida \& Makino 1993; Levison et al. 2010), which
slows their accretion rate significantly.
It has been proposed that
gas drag (Wetherill \& Stewart 1989) or mutual inelastic collisions
(Goldreich et al. 2004) prevent the dispersal of the planetesimals
by damping their orbital eccentricities, but in this case the cores
open gaps in the planetesimal disk (Levison \& Morbidelli 2007;
Levison et al. 2010), like the satellites Pan and Daphis open gaps in
Saturn's rings.  Thus the cores isolate themselves from the disk of
solids.  This effectively stops their growth.
It has been argued that
planet migration (Alibert et al. 2004) or the radial drift of sub-km
planetesimals due to gas drag (Rafikov 2004) break the isolation of
the cores from the disk of solids but, again, $N$-body simulations
show that the relative drift of planetesimals and cores simply
collects the former in resonances with the latter (Levison et al.
2010); this prevents the planetesimals from being accreted by the
cores.  In fact, only planetesimals smaller than a few tens of meters
drift in the disk fast enough to avoid trapping in any resonance with
a growing core (Weidenschilling \& David 1985). 

In a recent paper, Lambrechts \& Johansen (2012), hereafter LJ12, have
proposed a new model of core growth, which argues that, if the mass in
the disk is predominantly carried by pebbles of a few decimeters in
size, the largest planetesimals accrete pebbles very efficiently,
rapidly growing to several Earth masses (see also Johansen \& Lacerda
2010; Ormel \& Klahr 2010 and Murray-Clay et al. 2011).  More
specifically, this model builds on the recent planetesimal formation
model (Youdin \& Goodman 2005; Johansen et al. 2006, 2007, 2009) in
which large planetesimals (with sizes from $\sim 100$ up to
$\sim$1,000km) form by the collapse of a self-gravitating clump of
pebbles, concentrated to high densities by disk turbulence and the
streaming instability.  The mechanism by which, once formed,
planetesimals can keep accreting background pebbles is described
hereafter.

Pebbles are strongly coupled with the gas; thus
they encounter the already-formed planetesimals with a velocity
$\Delta v$ that is equal to the difference between the Keplerian
velocity and the orbital velocity of the gas (slightly sub-Keplerian
due to the outward pressure gradient). LJ12 define
the planetesimal {\it Bondi radius} as the distance at which the
planetesimal exerts a deflection of one radian on a
particle approaching with a velocity $\Delta v$:
\begin{equation}
R_B={{GM}\over{\Delta v^2}}
\label{Bondi}
\end{equation}
where $G$ is the gravitational constant and $M$ is the planetesimal
mass (obviously the deflection is larger if the particle passes closer
than $R_B$). LJ12 showed that all pebbles with a
stopping time $t_s$ smaller than the Bondi time $t_B=R_B/\Delta v$ that
pass within a distance $R=(t_s/t_B)^{1/2} R_B$ spiral down towards the
planetesimal and are accreted by it. Thus, the growth rate of the
planetesimal is:
\begin{equation}
\dot{M}=\pi\rho R^2\Delta v
\label{Bondi-accrete}
\end{equation}
where $\rho$ is the volume density of the pebbles in the disk.

From (\ref{Bondi}), the Bondi radius grows with the planetesimal
mass. LJ12 also showed that, when the Bondi radius
exceeds the scale height of the pebble layer, the accretion rate
becomes
\begin{equation}
\dot{M}=2R\Sigma \Delta v
\end{equation}
where $\Sigma$ is the surface density of the pebbles. Moreover, when the 
Bondi radius exceeds the Hill radius $R_H$, the accretion rate becomes
\begin{equation}
\dot{M}=2R_H \Sigma v_H
\label{Hill-accretion}
\end{equation}
where $v_H$ is the Hill velocity (i.e. the difference
in Keplerian velocities between two circular orbits separated by $R_H$).

With these formulae, and assuming that $\Sigma$ stays constant and is
close to the nominal density of solids in the MMSN, LJ12 showed that
the formation of 10 $M_E$ cores is possible within 1 My essentially
anywhere in the disk (up to $\sim 50 AU$).

There are two main advantages in the LJ12 model. First, it can form 10
$M_E$ cores of giant planets within the lifetime of the disk, a result
very difficult to achieve by other models. Second, because the
accretion rate (\ref{Bondi-accrete}) is very sensitive on the
planetesimal mass ($\dot{M}\propto M^2$), in practice only the largest
planetesimals formed in the turbulent model can effectively grow in
mass by this process: the minimal mass for triggering significant
Bondi accretion (see eq. \ref{Bondi-accrete}) is about the mass of
Ceres in the asteroid belt and about the mass of Pluto in the Kuiper
belt. Thus this model explains the maximal sizes observed in the
asteroid and Kuiper belt populations. In essence, in this model bodies
smaller than Ceres (respectively Pluto for the Kuiper belt) remained
small bodies (the asteroids and KBOs we see today), whereas those
bigger than this threshold kept accreting pebbles and became massive
objects (embryos) which then were removed by migrating away and
(possibly) participating to the build-up of the giant planets. Both
these aspects of the model are very appealing.

However the study conducted in LJ12, both in the
analytic and in the numerical parts, assumes that the motion of the gas
is not perturbed by the planetesimal. This assumption is good for a
Ceres-mass planetesimal, accreting as in (\ref{Bondi-accrete}), but
it is far from reality for planetary embryos (Earth mass or larger),
accreting through their Hill sphere as in (\ref{Hill-accretion}). In
fact, these objects modify the gas streamlines significantly: a
spiral density wave is formed in the disk and the gas near the orbit
of the planet has horseshoe motion. An over-density of gas is also
formed inside the planet's Hill sphere. It is not clear a priori what
are the effects of these structures on the pebble accretion rate. This
is precisely what we investigate in this paper with more realistic
hydro-dynamical simulations.

In section 2, we describe our methods: the simulation tool that we
have developed and the parameters that we adopt. In section 3 we
present our results, for two embryo masses and 4 pebble sizes. Our
goal is three-fold: (i) describe and understand the dynamics of the
pebbles for the different mass and size cases; (ii) evaluate the
accretion rate by the embryo and compare it with the LJ12 
estimate and (iii) evaluate the ``filtering factor'', that is the
fraction of the pebbles that do not drift by the orbit of the planet
because they are accreted by the embryo.  This factor is
important. If it is large, of a sequence of embryos radially
distributed in the disk, only the outermost one(s) can accrete;
instead, if it is small then the full system of embryos can
grow, in an oligarchic fashion.  Our conclusions and discussion of a
coherent scenario of giant planet formation conclude the paper in
section 4.

\section{Methods}

Our simulation software is based on the hydro-dynamical code FARGO,
developed to study planet-disk interactions in Masset (2000) and
publicly available at http://fargo.in2p3.fr/spip.php?auteur1.  In that
code, however, the N-body interaction among the bodies in the system
is studied with a Runge-Kutta integrator of 5th order. Because the
time-step of the integration is fixed, the Runge-Kutta algorithm is
not adequate to treat accurately the close encounters between objects
(e.g. the encounters between the pebbles and the planetary
embryo). Also, there is no prescription in the original code to detect
collisions.

Thus, we replaced the Runge-Kutta subroutine in FARGO with a 
code known as Symba. The latter is a variable-timestep symplectic code
developed in Duncan et al. (1998) to simulate quasi-Keplerian N-body
systems with mutual close encounters. Symba also identifies
collisions and merges the bodies that collide.  A technical difficulty
in interfacing FARGO with Symba was that the former is written in
C-language and the latter in Fortran. This required extensive
modifications of several subroutines of the FARGO package.

In our simulations, the embryo is a massive object: it influences the
evolution of the gas but, for simplicity, we cancel the influence of
the gas disk on the embryo, so that the latter remains on a fixed,
non-migrating orbit. This approximation is reasonable, 
as long as the migration of the embryo is slow
compared to the radial drift of the pebbles due to gas drag.

In the FARGO code, the gas is discretized over a polar two-dimensional
grid. {However, in our modified code, the pebbles can evolve in a
  three-dimensional space. The gas drag on the pebbles is then
  computed as follows.  We consider a spherical coordinate system
  $r,\theta,\phi$, where $r,\theta$ are the radial and azimuthal
coordinates on the disk mid-plane.  At each timestep we identify the
disk grid cell where the spherical projection of the pebble falls on the midplane. This
depends on the $r,\theta$ coordinates of the pebble}.  We consider
the values of the gas surface density $\Sigma$ and radial and
tangential velocities $v_r, v_\theta$ that the gas has in that cell as
well as in its 8 neighboring cells. We interpolate the set of 9 values
for each field with a polynomial function of type
$(a+bx+cx^2)+(a'+b'x+c'x^2)y+(a''+b''x+c''x^2)y^2$ and we use it to
evaluate $\Sigma, v_r$ and $v_\theta$ of the gas at the $r,\theta$
location of the pebble. Moreover, we assume that the vertical velocity
of the gas $v_z$ is zero. This sets the local relative velocity
$\vec{\delta v}$ of the gas and the pebble.  The local volume density
of the gas at the pebble location is computed as
$\rho=\Sigma/(\sqrt{\pi} H) \exp(-z^2/H^2)$ where $z$ is the vertical
coordinate of the pebble and $H/r$ is the assumed aspect ratio of the
disk, given as an input of the simulation. Finally, the drag suffered
by the pebble is computed as $\dot{\vec{v}}=-1\vec{\delta v}/t_s$,
where $t_s$ is the stopping time. The stopping time is also evaluated
locally following Adachi et al. (1976), given the size and bulk
density of the pebble (given as an input of the simulation) and the
local volume density of the gas and Mach \& Knudsen numbers.

\subsection{Simulation parameters}

Our simulations follow the dynamics of one embryo and a set of pebbles.  We
consider planetary embryos of 1 or 5 $M_E$.  They are placed on a
circular non-migrating orbit at 0.8 AU.  The disk's surface density is
equal to 1800g/cm$^2$ at 1 AU, corresponding to the MMSN, with a
radial profile proportional to $1/r$. The aspect ratio is 5.6\%. We
follow an $\alpha$-prescription for the disk's viscosity (Shakura \&
Sunyaev 1973), with $\alpha=6\times 10^{-3}$. The parameters above
constitute just one set of (typical) disk parameters. Different disk
parameters would affect mainly the pebble stopping
time. Given that we will study the dynamics of pebbles over a wide
range of sizes, our exploration of the stopping-time parameter space
will be exhaustive even if working with just one disk model.

The disk is modeled over a grid extended from 0.75 to 1.12 AU, with a
resolution of 160 concentric rings and 720 sectors. As the disk radial
extension is narrow, we use non-reflecting boundary
conditions, so that the spiral density wave launched by the embryo
does not make the surface density of the disk rough with multiple
reflections.

\begin{figure}[t!] 
\centerline{\includegraphics[width=10.cm]{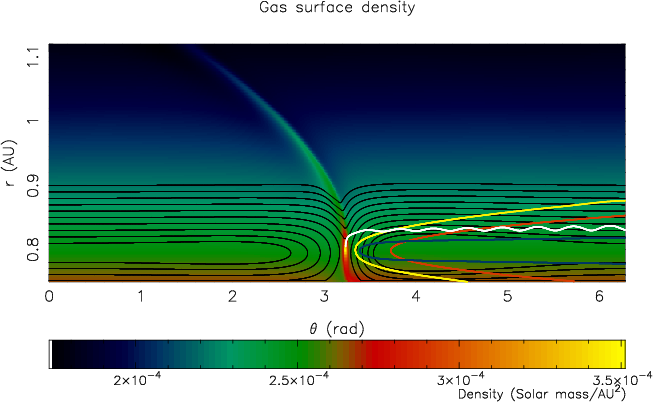}}
\caption{\small Color map of the surface density of the gas in a disk
  in the presence of a $5 M_E$ embryo at $r=0.8, \theta=3.22$. The
  black curves show some gas streamlines and the blue, red, yellow and
  white curves depict the trajectories of 5cm, 20cm, 1m and 10m
  pebbles, respectively, in the frame corotating with the embryo.}
\label{Surf}
\end{figure}

An image of the surface density of the disk is shown in
Fig.\ref{Surf}, with a color scale. The black curves show 
streamlines of the gas in the reference frame corotating with the
embryo. The position of the embryo is highlighted by a
local maximum of the disk's surface density. Notice that the spiral density
wave (the slanted over-density structure departing from the planet)
does not have any reflection at the border of the frame. Therefore, it
is not necessary to simulate a wider radial portion of the disk. 

The gravitational interaction between the embryo and the gas is
smoothed as $1/(\Delta+\epsilon)^2$ where $\Delta$ is the distance
between the embryo and a gas fluid element and $\epsilon$ is a
smoothing parameter. We adopt the quite standard choice $\epsilon=0.6
R_H$.

We simulated pebbles of 5cm, 20cm, 1m and 10m in radius, with a bulk
density of 1g/cm$^3$.  With our disk model this corresponds to the
following stopping times at 1 AU: 0.017, 0.22, 2.78, 74.07 $cu$
respectively, {where $cu$ is the time code unit, which is $1/2\pi$ of
an year}. Even if some of these particles are rather boulder-sized, we
still call them pebbles, for simplicity. The pebbles are assumed to
orbit on the same plane as the disk and the embryo {so that our
  simulations become effectively two-dimensional}. A trajectory of one
pebble of each size is shown in Fig.~\ref{Surf} and described in the
next section.  The timestep used in the integration is the minimum
between 1/4 of the particle stopping time and 0.1 $cu$. The latter is the
timestep $dt$ imposed by the CFL condition (Courant et al. 1928) for
the numerical solution of the hydrodynamical equations, given the
resolution and the extension of our grid.  Remember that the Symba
algorithm effectively reduces the timestep for the integration of the
particles approaching the embryo, through a clever subdivision of the
embryo's gravitational potential (Duncan et al. 1998).

To measure the accretion rate and the filtering factor (see last
paragraph of Sect.~1) we proceed in two steps.  We first integrate one
pebble starting from the outer boundary of the disk at $r=1.12$~AU at
opposition with respect to the embryo. As the pebble migrates inward,
we record the sequence of heliocentric distance values $r_k$ that the
pebble has at subsequent oppositions. We denote by $r_N$ the smallest
of the $r_k$ values with $r_K>0.8$~AU, i.e. the heliocentric distance
at the last opposition before crossing the orbit of the
embryo\footnote{For the 10m ``pebble'', we chose instead $N=2$, so
  that it is still far enough from the embryo to preserve a
  quasi-circular orbit.}. Then, we simulate a large number of pebbles
(usually 100, but this number can change from case to case to achieve
an appropriate resolution), initially at opposition and with
heliocentric distances uniformly distributed in the range $[r_N,
  r_{N-1}]$. Of this beam of pebbles, we measure the fraction $F$ that
are accreted by the planet.  $F$ is the ``filtering factor'' defined
in the introduction, while $1-F$ is the fraction of the beam that
drift across the orbit of the embryo.  The accretion rate is then:
\begin{equation}
\dot M= \Sigma (r_{N-1}-r_N) \times F \Delta v
\label{our-accretion}
\end{equation}
where $\Delta v$ is the difference in orbital velocity between the
embryo and a pebble placed in the middle of the $[r_N, r_{N-1}]$
interval and $\Sigma$ is the surface density of the disk of
pebbles. Thus, our result is parametric in $\Sigma$ and can be directly
compared with the expression (\ref{Hill-accretion}) from LJ12.
 
\section{Results}

Fig.~\ref{Surf} shows the trajectory of one pebble for each of the 4
sizes that we consider in this study.  The evolution is shown in a
rotating reference frame, with the embryo fixed at $r=0.8$~AU and
$\theta=3.22$. Only the last synodic revolution is shown, i.e. that
leading to the crossing of the orbit of the embryo. The trajectories
of the 5cm, 20cm, 1m and 10m pebbles are depicted in blue, red, yellow
and white, respectively. The mass of the embryo is $5 M_E$.

Several considerations should be made, by looking at these
trajectories.  First, notice that the stopping time (which increases
with particle size) is not simply related to the migration speed. The
particles with the fastest radial migrations are those with radii of
20cm and 1m (with migration rates roughly equal to each other). This
is well known (Weidenschilling 1977b).  The 5cm particle migrates more
slowly because it is more coupled with the gas (which does not have
any net radial motion); the 10m particle migrates more slowly than the
1m particles because it is less sensitive to gas drag.

The U-turns drawn by the trajectories of the 20cm and 1m pebbles are not
related to horseshoe dynamics. They are simply due to the fact that the
angular velocity of a Keplerian orbit decreases as $1/r^{3/2}$ so
that, in a reference system corotating with an object at a fixed
distance $r_0$ (like Fig.~\ref{Surf} corotating with the embryo at
$r_0=0.8$AU) a particle moves from the right to the left if it has
$r>r_0$ and from the left to the right if it has $r<r_0$. Thus, as the
particle drifts from $r>r_0$ to $r<r_0$ it has to make a U-turn. In
other terms, these bended trajectories are not due to the
gravitational effect of the embryo but just to the rigid rotation of
the reference frame. Only particles passing close to the embryo
(shown below) are affected by the gravitational attraction of the latter.

Instead, the U-turn of the 5cm pebble is due to horseshoe
dynamics. This is apparent from the fact that the radial motion is
fast only in the vicinity of the embryo. The particle shown in
Fig.~\ref{Surf} traces approximately the maximal width of a particle horseshoe
trajectory. Notice that the width of the horseshoe region for the
particles is much narrower than that for the gas, highlighted by the
U-turning streamlines.

Finally, the trajectory of the 10m object shows radial oscillations as
it approaches towards the embryo. This is due to its orbital
eccentricity, acquired during previous passages at conjunction with
the embryo, and not fully damped by the gas drag in half a synodic period
(i.e. from conjunction to opposition).  This is not the case from the
other particles for two reasons: gas drag is stronger and erases the
eccentricity faster (e.g. the 5cm pebbles) and/or 
given their fast radial drift the
pebbles passed too far from the embryo at the previous
conjunction to acquire a large eccentricity (e.g. the 20cm and 1m particles). 

We now focus on the dynamics in the very vicinity of the embryo, leading to 
pebble accretion and the embryo's growth.

\begin{figure}[t!] 
\centerline{\includegraphics[width=9.cm]{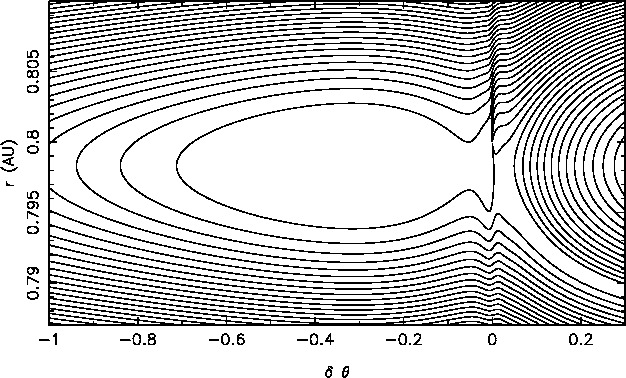}}
\caption{\small the trajectories of 5cm pebbles approaching a 1 $M_E$ embryo in relative $r,\theta$ coordinates.}
\label{5cm}
\end{figure}

Fig.~\ref{5cm} shows the trajectories of several 5cm pebbles as they
pass close to the $1 M_E$ embryo. Each line is a different
particle. The $x$-axis reports the difference $\delta\theta$ between
the $\theta$ values of the pebble and the embryo. Notice on the right
hand side of the figure the particles doing the horseshoe U-turn
(those which always have $\delta\theta>0$). Instead, the particles that
are initially in circulation, i.e. that have $r>0.8$~AU and
$\delta\theta$ passing from positive to negative, eventually cross the
embryo's orbit far behind the embryo itself, when $\delta\theta <
-0.6$.  Notice that the U-turn of the pebbles occurs at $r\sim 0.798$,
i..e slightly inside the embryo's orbit ($r=0.8$). This is because
these pebbles are strongly coupled with the gas, which has a
sub-Keplerian orbital velocity; thus the exact corotation radius is
shifted towards the Sun. In total, 7 pebbles collide with the embryo,
assumed to have the Earth's radius, as one can see from the
trajectories diving towards the point with coordinate $(0, 0.8)$. Six
of these pebbles collide with the embryo in the approach phase, when
$\delta\theta$ is evolving from positive to negative. However, one
pebble hits the embryo ``from the back'', i.e. after having crossed
the embryo's orbit and with $\delta\theta$ evolving from negative to
positive.

\begin{figure}[t!] 
\centerline{\includegraphics[width=9.cm]{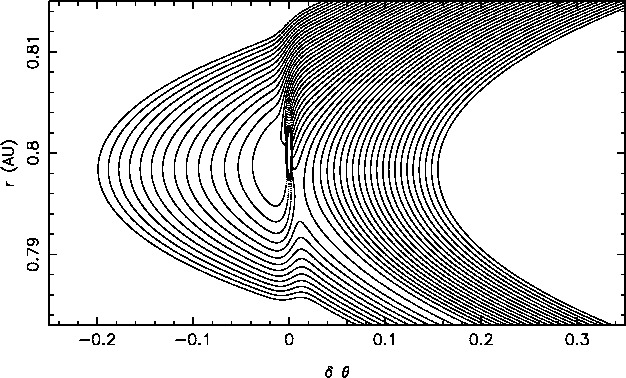}}
\caption{\small The same as Fig.~\ref{5cm} but for 20cm pebbles.}
\label{20cm}
\end{figure}

Fig.~\ref{20cm} is the same but for 20cm pebbles. The figure is
qualitatively similar to the previous one, but the trajectories are much
more inclined due to the fact that these pebbles drift much faster
towards the Sun (beware that the scales in Fig.~\ref{5cm}
and~\ref{20cm} are different) and, consequently, the particles that
cross the orbit of the embryo at negative $\delta\theta$ do so closer
to the embryo (with $\delta\theta>-0.2$ instead of $<-0.6$).  The
thick ellipse drawn around $(0,0.8)$ marks a circle centered on the
embryo with a radius equal to 1/3 of its Hill radius. The resolution
of the output of the simulation is too coarse to resolve the
trajectories inside this radius. However, the internal timestep of the
Symba algorithm is good enough to resolve the physical collision of
the pebbles with the embryo. There are 18 pebbles entering inside the
circle, 4 of which enter ``from the back'', i.e. after having crossed the
embryo's orbit. Of these 18 pebbles, 13 physically collide with the
embryo.  The remaining ones are slingshot away by the embryo during the
close encounter and exit the domain covered by our grid. 

\begin{figure}[t!] 
\centerline{\includegraphics[width=9.cm]{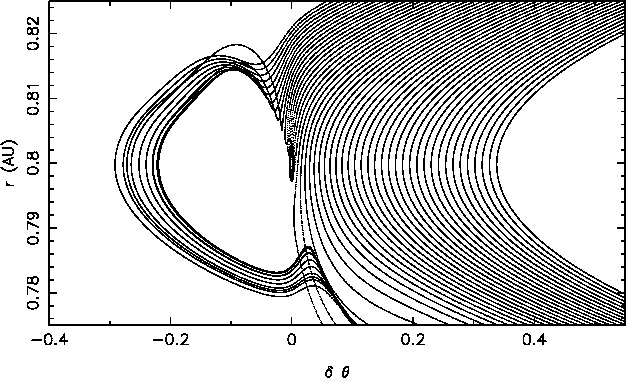}}
\caption{\small The same as Fig.~\ref{5cm} but for 1m particles.}
\label{1m}
\end{figure}

Fig.~\ref{1m} is for one-meter particles. The major difference with
respect to the previous figure is the strong kink that particles
receive when passing from positive to negative $\delta\theta$. This
happens because these particles are less coupled to the gas and
therefore they can be scattered to large eccentricity when they pass
in conjunction with the embryo. Consequently, the 1-m particles pass
much further away from the embryo than the 20cm-particles when they
cross the embryo's orbit, and therefore they cannot collide with the
embryo ``from the back''.

\begin{figure}[t!] 
\centerline{\includegraphics[width=9.cm]{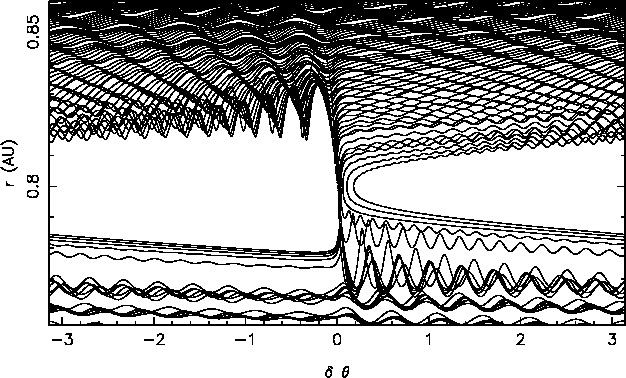}}
\caption{\small The same as Fig.~\ref{5cm} but for 10m particles.}
\label{10m1Me}
\end{figure}

\begin{figure}[t!] 
\centerline{\includegraphics[width=9.cm]{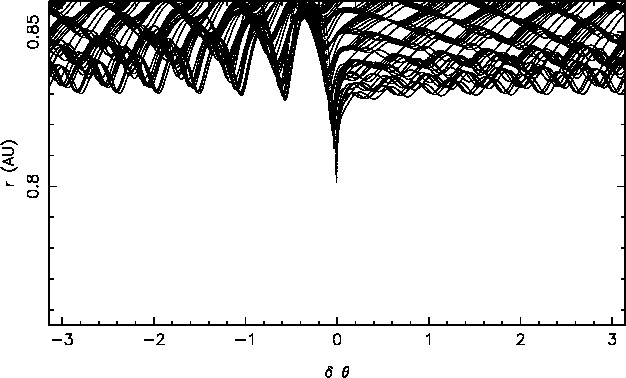}}
\caption{\small The same as Fig.~\ref{10m1Me} but for a $5 M_E$ embryo.}
\label{10m5Me}
\end{figure}

Fig.~\ref{10m1Me} is for 10m-particles. For these particles we change
the scale of the plot, showing the whole disk up to $r=0.86$~AU. This is
because gas drag is weak, and the particles conserve part of the
eccentricity that they acquire during the previous conjunction with
the embryo for a full synodic period. Consequently, their dynamical
behavior when crossing the embryo's orbit inherits the dynamical
history recorded over several synodic revolutions. Because the radial
drift during a synodic revolution is small, all particles have to pass
eventually very close to the embryo, so that many of them (40\%)
are accreted. The particles that safely cross the embryo's orbit do so
with a horseshoe U-turn at $\delta\theta>0$. If the embryo is more massive 
the fraction of accreted particles increases and, for a 5~$M_E$ embryo, 
the accretion efficiency is 100\% (see Fig.~\ref{10m5Me}). 

We now come to a synthesis of the results for what concerns the embryo's mass 
accretion rate and filtering factor. 

\begin{figure}[t!] 
\centerline{\includegraphics[width=9.cm]{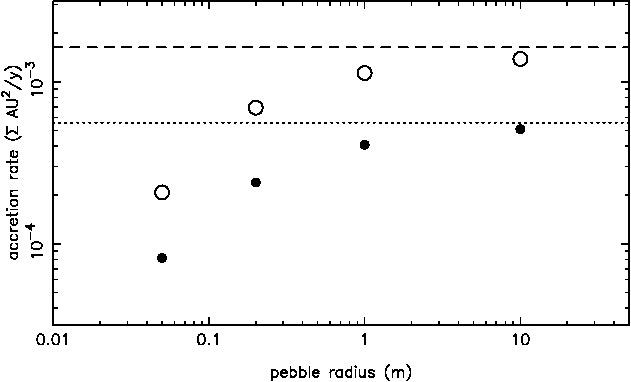}}
\caption{\small The accretion rate as a function of pebble size for
  embryos of $1 M_E$ (filled dots) and $5 M_E$ (open dots)
  respectively. The dotted and dashed lines indicate the accretion
  rates estimated in LJ12 for these two
  embryo masses.}
\label{accretion}
\end{figure}

Fig.~\ref{accretion} shows the mass accretion rate for the embryos of
masses $1 M_E$ (filled dots) and $5 M_E$ (open dots) respectively, as
a function of the pebbles size. The mass accretion rate is normalized
by $\Sigma$ (the surface density of the pebbles).  The dotted and
dashed horizontal lines show the accretion rate estimated by LJ12, for
embryos of masses $1 M_E$ and $5 M_E$ respectively. As one can see,
our numerically measured accretion rates are within an order of
magnitude of the estimated rates, and tend asymptotically to the
estimated values for growing pebble sizes. The drop of the accretion
rate with particle size is due mainly because a larger fraction of the
particles that enter the Hill sphere of the embryo are dragged away by
the flow of the gas, instead of falling onto the central object.  This
phenomenon was expected in LJ12, who predicted that their accretion
rate is valid for pebbles with stopping time of the order 0.1--1.
Remember that 20cm pebbles in our simulation have $t_s=0.22$.  Thus,
our results suggest that the nominal accretion rate of LJ12 applies
instead from $t_s\sim 1$; however, it remains valid up to stopping
times larger than predicted, i.e. up to at least $\sim 100$ (the value
of $t_s$ for the 10m particles is 75).

The scaling relative to the embryo's mass
predicted by LJ12 ($\dot M\propto R_h v_h$,
i.e. $\dot M\propto M^{2/3}$) is confirmed by our numerical
simulations for all tested pebble sizes.

\begin{figure}[t!] 
\centerline{\includegraphics[width=9.cm]{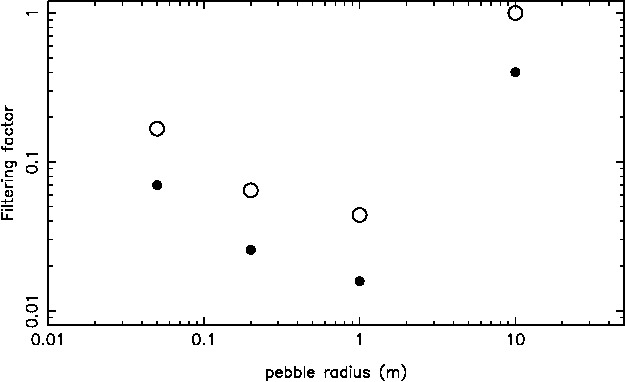}}
\caption{\small The filtering factor as a function of pebble size for
  embryos of $1 M_E$ (filled dots) and $5 M_E$ (open dots)
  respectively.}
\label{filtering}
\end{figure}

Figure~\ref{filtering} shows the filtering factor as a function of
pebble size, again for $5 M_E$ (open dots) and $1 M_E$ (filled dots)
embryos. The filling factor scales again as
$M^{2/3}$.  The filtering factor decreases with increasing drift speed
and therefore it is the smallest for the m-sized particles. Except for
particles of multiple meters, the filtering factor is small (a few
percent to 10\%), which implies that a system of numerous embryos
(with a number of objects roughly proportional to $1/F$) can grow
simultaneously in the disk. Notice that the embryos should grow in an
oligarchic fashion, due to the $M^{2/3}$ dependence of the accretion
rate. Instead, if the disk is dominated by multi-meter particles, only
a few embryos can grow, with eventually the outermost one capturing
all the material.

\begin{figure}[t!] 
\centerline{\includegraphics[width=9.cm]{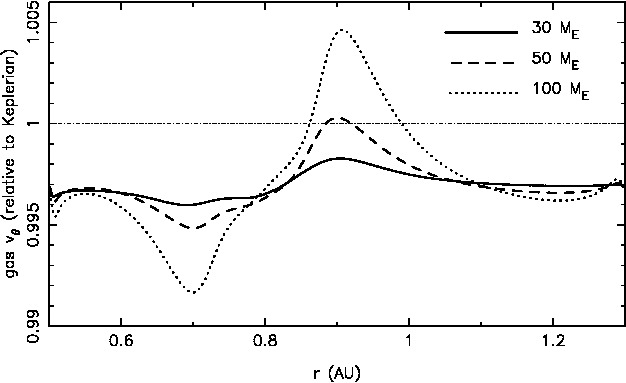}}
\caption{\small The gas azimuthal velocity relative to the Keplerian
  value, as a function of orbital radius, for disks with embedded
  planets of 30, 50 and 100 Earth masses. The horizontal line marks
  the boundary between sub-Keplerian and super-Keplerian rotation.}
\label{rotation}
\end{figure}

In the LJ12 model, there is no upper limit in mass for an accreting
embryo, as long as pebbles are available: an embryo keeps accreting at
a rate proportional to $M^{2/3}\Sigma$. In reality, however, if the
total mass of the planet (the mass of the solid embryo plus the mass
of the gas eventually accreted in its atmosphere) becomes large
enough, the surface density of the gas distribution is modified. The
neighborhood of the planet orbit becomes partially depleted of gas (a
shallow gap is opened) and this changes the pressure gradient in the
disk and the angular velocity of the gas. Eventually the gas becomes
super-Keplerian at some location outside of the planet's orbit and
this stops the inward radial drift of pebbles and small
planetesimals. The accretion of the planet by the LJ12 mechanism
suddenly stops. Fig.\ref{rotation} shows the ratio between the
azimuthal velocity of the gas and that of a Keplerian circular orbit
as a function of radius, for planet masses of 30, 50 and 100 $M_E$
placed at $r=0.8$. For these tests, we have increased the
  disk's radial extension to the interval 0.5--1.3 AU. For the disk
parameters that we chose (see sect. 2.1), we find that the disk
becomes locally super-Keplerian is the planet's mass is $\sim 50
M_E$. This mass would be reduced if the disk had a lower viscosity
and/or scale height. Whatever the exact value of the limit mass of the
planet, our result shows that the LJ12 mechanism is capable of feeding
a solids to a giant planet until its mass is several tens of Earth
masses.

This result has two main implications.  First, the runaway accretion
of the atmosphere onto the embryo might start earlier than in the
classic Pollack et al. (1996) model. In fact, in Pollack et al. the
embryo's accretion stalls when the mass of the embryo is about
10~$M_E$, because the feeding zone for solids is strongly
depleted. Then,the embryo's accretion of both gas and solids continues
very slowly, with the consequence that the critical mass for
triggering runaway capture of gas ($\sim 30 M_E$) is reached only
after several My. In the LJ12 scenario, the mass accretion of the
embryo would not slow down, so that the runaway accretion of gas, in
principle, might start at an earlier time.  We notice however that a
high accretion rate of solids has also a negative effect on gas
accretion because it releases a lot of energy on the embryo which
needs to be evacuated (Dodson-Robinson \& Bodenheimer 2010); thus,
the net effect of pebble-accretion on gas-accretion will need to be
investigated in hydrodynamical simulations accounting for heat
transfer.
Second, the accretion of pebbles
until the planet mass achieves $\sim 50$ $M_E$ can help explaining
the large ratio between heavy elements and hydrogen and helium in the
atmosphere of Jupiter, which is 3-4 times solar (Wong et al.
2004). Enriching the giant planets in heavy elements is not
straightforward (Guillot \& Gladman 2000) by other mechanisms; thus
this is another appealing aspect of the LJ12 model.

\section{Conclusions}

In this paper we have tested the scenario of giant planet core
formation proposed by LJ12 with
hydrodynamical simulations that fully account for (i) the interaction
between the growing core and the gas of the disk and (ii) the local drag
exerted by the gas on pebbles and boulders.

We have found that the pebble dynamics in the vicinity of the
planetary embryo is non-trivial, and that it changes significantly
with the pebble size. Nevertheless, the accretion rate of the embryo
that we measure is within an order of magnitude of the rate estimated in 
LJ12 and tends to their value with increasing pebble-size. 

The accretion of pebbles can continue until the embryo's mass is of
the order of 50 Earth masses (in solids and gas). This may have
important implications on the onset of runaway accretion of gas by the
growing giant planets and can help explaining the enrichment in solids
observed in Jupiter's atmosphere.

The actual accretion rate of an embryo depends on the amount of mass
$\Sigma$ available in pebbles, which is not known a priori, given that
pebbles are consumed by the formation of planetesimals and by the
accretion of the embryos themselves. Nevertheless, the accretion rates
that we find in Fig.~\ref{accretion} are potentially large. For
instance, if a MMSN of solids (20g/cm$^2$ at 1 AU) were available in
20cm pebbles, the mass doubling time for a 1~$M_E$ embryo would be
only 5,500 years! This illustrates the importance of the LJ12 model
for the growth of giant planets cores.

Given that the LJ12 model is built in the same framework as the model
that explains the rapid formation of planetesimals (Johansen et al.
2007), we believe that the community has now, for the first time, a
coherent scenario to explain the the early phases of planet
growth. However, we do not see any evident reason for which only a
small number (4-6) of giant planet cores should grow in the disk, as
suggested by the number of giant planets of our solar system,
including rogue ice-giants potentially lost during the dynamical
evolution that followed planet formation (Nesvorny 2011; Nesvorny \&
Morbidelli 2012). Thus, we think that, most likely, the LJ12 model
explains the formation of massive planetary embryos of a few Earth
masses, but an additional stage is needed to form the giant planet
cores (10 Earth masses or more).

This is the scenario that we envision. The embryos formed by the
LJ12 mechanism, once massive enough, start to
migrate in the disk due to planet-disk interactions. Recent results on
migration in radiatively cooling disks (Paardekooper \& Mallema
2006; Baruteau \& Masset 2008; Kley \& Crida 2008; Paardekooper et
al. 2010; Masset \& Casoli 2010; Bitsch \& Kley 2011) show that
the embryos migrate from all directions toward an orbital radius where
migration is canceled out by the compensation of competing
torques. This convergent migration towards the same region can promote
the mutual accretion of embryos, eventually reducing a system of a
large number of embryos into a system of a smaller number of larger
objects, i.e. a handful of giant planet cores (Horn et al. 2012). Admittedly,
this scenario is still speculative and more work is needed to prove
its validity.  We stress, however, that a final phase of core
formation characterized by mutual collisions of embryos would explain,
in a natural way, the massive impacts that are needed to explain the
current obliquities of Uranus and Neptune (Morbidelli et al. 2012).

\section{Acknowledgments}
Alessandro Morbidelli thanks Germany’s Helmholtz Alliance for
providing support through its “Planetary Evolution and Life”
programme. David Nesvorny thanks the Observatoire de la C\^ote d'Azur
for hospitality during his sabbatical in Nice. Both authors thank an
anonymous reviewer for reading the manuscript in details.

\section{References}

\begin{itemize}
\item[--] Adachi, I., Hayashi, C., \&
Nakazawa, K.\ 1976.\ Prog. Theor. 
Phys. 56, 1756 
\item[--] Alibert, Y., Mordasini, C., \& Benz, W.\ 2004.\ A\&A 417, L25. 
\item[--] Baruteau, C., \& 
Masset, F.\ 2008.\ ApJ 672, 1054. 
\item[--] Bitsch, B., \& Kley, W.\ 2011.\ A\&A 536, A77. 
\item[--] Chambers, J.\ 2006.\ Icarus 180, 496. 
\item[--] Courant, R., Friedrichs, K., \& Lewy, H. 1928. Mathematische Annalen 100, 32.
\item[--] Dodson-Robinson, S.~E., \& Bodenheimer, P.\ 2010.\ 
Icarus 207, 491. 
\item[--] Duncan, M.~J., Levison, 
H.~F., \& Lee, M.~H.\ 1998.\ AJ 116, 2067. 
\item[--] Greenberg, R., 
Hartmann, W.~K., Chapman, C.~R., \& Wacker, J.~F.\ 1978.\ Icarus 35, 1. 
\item[--] Goldreich, P., 
Lithwick, Y., \& Sari, R.\ 2004.\ ApJ 614, 49. 
\item[--] Guillot, T., \&
Gladman, B.\ 2000.\ Disks, Planetesimals, and Planets 219, 475. 
\item[--] Haisch, K.~E., 
Lada, E.~A., \& Lada, C.~J.\ 2001.\ ApJ 553, L153. 
\item[--] Hayashi, C.\ 1981. Prog. Theor.
Phys. Suppl., 70, 35.
\item[--] Horn, B., Lyra, W., Mac 
Low, M.-M., \& S{\'a}ndor, Z.\ 2012.\ ApJ 750, 34. 
\item[--] Ida, S., \& Makino, J.\ 
1993.\ Icarus 106, 210. 
\item[--] Johansen, A., Klahr, \&
H., Henning, T.\ 2006.\ ApJ 636, 1121. 
\item[--] Johansen, A., Oishi, 
J.~S., Mac Low, M.-M., Klahr, H., Henning, T., \& Youdin, A.\ 2007.\  Nature 448, 
1022. 
\item[--] Johansen, A., Youdin, 
A., \& Mac Low, M.-M.\ 2009.\  ApJ 704, L75-L79. 
\item[--] Johansen, A., \& Lacerda, P.\ 2010, MNRAS, 404, 475 
\item[--] Kley, W., \& Crida, A.\ 2008.\ A\&A 487, L9. 
\item[--] Kokubo, E., \& Ida, S.\ 
1998.\ Icarus 131, 171. 
\item[--] Lambrechts, M. \& Johansen, A., 2012. 
A\&A 544, A32.
\item[--] Levison, H.~F., \&
Morbidelli, A.\ 2007.\ Icarus 189, 196. 
\item[--] Levison, H.~F., 
Thommes, E., \& Duncan, M.~J.\ 2010.\ AJ 139, 
1297. 
\item[--]  Masset, F.\ 2000.\ A\&A Suppl. Series 141, 165. 
\item[--] Masset, F.~S., \&
Casoli, J.\ 2010.\ ApJ 723, 1393. 
\item[--]  Morbidelli, A., 
Tsiganis, K., Batygin, K., Crida, A., \& Gomes, R.\ 2012.\ Icarus 219, 737. 
\item[--]  Murray-Clay, R., 
Kratter, K., \& Youdin, A.\ 2011.\ AAS/Division for Extreme Solar Systems Abstracts 2, 
804. 
\item[--]  Nesvorn{\'y}, D.\ 2011.\ 
ApJ 742,  L22. 
\item[--]  Nesvorn{\'y}, D. \& Morbidelli A., 2012. AJ, in press. 
\item[--] Ormel, C.~W., \& Klahr, H.~H.\ 2010, A\&A, 520, A43 
\item[--] Paardekooper, S.-J., \& Mellema, G.\ 2006.\ A\&A 459, L17. 
\item[--] Paardekooper, 
S.-J., Baruteau, C., Crida, A., \& Kley, W.\ 2010.\ MNRAS 401, 1950. 
\item[--] Podolak, M., \& Zucker, S.\ 2004.\ MAPS 39, 1859. 
\item[--] Pollack, J.~B., 
Hubickyj, O., Bodenheimer, P., Lissauer, J.~J., Podolak, M., \& Greenzweig, 
Y.\ 1996.\ Icarus 124, 62. 
\item[--] Rafikov, R.~R.\ 2004.\ ApJ 128, 1348. 
\item[--] Shakura, N.~I., \& 
Sunyaev, R.~A.\ 1973.\ A\&A 24, 337-355. 
\item[--] Thommes, E.~W., Duncan, 
M.~J., \& Levison, H.~F.\ 2003.\ Icarus 
161, 431. 
\item[--]  Youdin, A.~N., \& 
Goodman, J.\ 2005.\ ApJ 620, 459. 
\item[--] Weidenschilling, S.~J.\ 1977.\ ApSS 51, 153. 
\item[--] Weidenschilling, 
S.~J.\ 1977b.\ MNRAS 180, 57. 
\item[--] Weidenschilling, S.~J., \& Davis, D.~R.\ 1985.\ Icarus 62, 16. 
\item[--] Wetherill, 
G.~W., \& Stewart, G.~R.\ 1989.\ Icarus 77, 330-357. 
\end{itemize}

\end{document}